\documentclass[aps,prl,preprint,showpacs,groupedaddress]{revtex4}
\usepackage {amssymb}
\usepackage {amsmath}
\usepackage {graphicx}
\usepackage {longtable}

\begin{document}
\title{Induced radiation processes in single-bubble sonoluminescence}

\author{Fedor V.Prigara}
\affiliation{Institute of Microelectronics and Informatics,
Russian Academy of Sciences,\\ 21 Universitetskaya, Yaroslavl
150007, Russia} \email{fprigara@imras.yar.ru}

\date{\today}

\begin{abstract}

According to the recent revision of the theory of thermal
radiation, thermal black-body radiation has an induced origin. We
show that in single-bubble sonoluminescence thermal radiation is
emitted by a spherical resonator, coincident with the
sonoluminescing bubble itself, instead of the ensemble of
elementary resonators emitting thermal black-body radiation in the
case of open gaseous media. For a given wavelength, the diameter
of the resonator is fixed, and this explains the very high
constancy in phase of light flashes from the sonoluminesing
bubble, which is better than the constancy of period of a driving
acoustic wave.

\end{abstract}

\pacs{78.60.Mq, 42.65.Re, 47.40.-x}

\maketitle

Sonoluminescence is the emission of light flashes in continuum
from an oscillating micron-size bubble at a pressure anti-node of
a sound wave in a liquid [1-4]. Sonoluminescence has been
discovered as early as in 1930s and remained so far an enigmatic
phenomenon. Recent computation of the spectrum intensities from an
optically thin model of emitting gas in a sonoluminescing bubble
[5] have shown the great discrepancies between the calculated
values and experimental results for a single-bubble
sonoluminescence. It was shown, however, recently [6, 7] that
thermal black-body radiation has an induced origin and the
emitting gas is automatically thick with respect to the emission
process. Here we show that the account for stimulated origin of
thermal radiation is required to explain the observed properties
of single-bubble sonoluminescence.

The induced origin of thermal radio emission follows from the relations
between Einstein's coefficients for a spontaneous and induced emission of
radiation [6]. The strong argument in a favor of an induced origin of
thermal black-body radiation is that the spectral energy density in the
whole range of spectrum is described by a single Planck's function. So if
thermal radio emission is stimulated, then thermal radiation in other
spectral regions also should have the induced character.

According to this conception, thermal emission from non-uniform gas is
produced by an ensemble of individual emitters. Each of these emitters is an
elementary resonator the size of which has an order of magnitude of mean
free path , \textit{l}, of photons,

\begin{equation}
\label{eq1}
l = 1/\left( {n\sigma}  \right),
\end{equation}

\noindent
where \textit{n} is the number density of particles and $\sigma $ is the
absorption cross-section. The emission of each elementary resonator is
coherent, with the wavelength

\begin{equation}
\label{eq2}
\lambda = al,
\end{equation}

\noindent
where \textit{a} is a dimensionless constant, and thermal emission of
gaseous layer is incoherent sum of radiation produced by individual
emitters. An elementary resonator emits in the direction opposite to the
direction of the density gradient. The wall of the resonator corresponding
to the lower density is half-transparent due to the decrease of absorption
with the decreasing gas density.

In the case of single-bubble sonoluminescence, an elementary
emitter of thermal black-body radiation is a spherical resonator,
coincident with the sonoluminescing bubble itself, instead of the
ensemble of elementary resonators characteristic for open
non-uniform gaseous media. For a spherical electromagnetic wave in
the resonator, the electric field strength in the wave is given by
the formula

\begin{equation}
\label{eq3}
E = Asin\left( {kr} \right)/r,
\end{equation}

\noindent
where \textit{A} is a constant, $k = 2\pi n_{r} /\lambda $ is the wave
number, $\lambda $ is the wavelength in vacuum, $n_{r} $ is the refraction
index, and \textit{r} is the radial coordinate. At the surface of the
spherical resonator, we have a node, so

\begin{equation}
\label{eq4}
R = \lambda /\left( {2n_{r}}  \right),
\end{equation}

\noindent
where \textit{R} is the radius of the bubble.

It is clear from the last equation that, for a given wavelength $\lambda $,
the radius of the bubble at which the radiation with the wavelength $\lambda
$ is emitted is fixed. This explains the very high constancy in phase of
sonoluminescence flashes with respect to the period of driving sound wave,
exceeding the constancy of period of the acoustic wave itself [2-4].

The measured emission duration [8] seems to correspond to narrow spectral
bands and is determined in fact by the bandwidth. The full duration of the
sonoluminescence flash in continuum is given by the formula

\begin{equation}
\label{eq5}
\tau = \Delta \lambda /\left( {2n_{r} v} \right),
\end{equation}

\noindent
where $\Delta \lambda = \lambda _{max} - \lambda _{min} $ is the difference
between the maximum and minimum values of the wavelength respectively, and
\textit{v} is the velocity of the implosion shock wave at the time of light
emission.

The initial velocity of the bubble imploding is equal to the velocity of
sound in the liquid. During the implosion process, the velocity of the
implosion shock wave increases [2]. The kinetic energy of particles in the
shock wave is

\begin{equation}
\label{eq6}
\varepsilon = mv^{2}/2,
\end{equation}

\noindent
where \textit{m} is the mass of a molecule, and can achieve the value of
about $1eV$ at the time of light emission. This kinetic energy converts into
the thermal energy of a gas in the bubble by generation of plasma waves at
sufficient small radii of the bubble when the density and ionization degree
of the gas in the bubble are sufficiently large.

The addition of small amounts of noble gases, such as helium, argon, or
xenon, to the gas in the sonoluminescing bubble increases the intensity of
the emitted light dramatically [9]. This effect seems to be analogous to the
argon catalysis of the excessive Balmer line broadening in gas discharges
[10]. Since the emission is produced by a single resonator, it can be
amplified by a laser mechanism. The argon doping seems to play an important
role in the creation of the inversion of the energy level population which
is required for such an amplification.

\begin{center}
---------------------------------------------------------------
\end{center}

[1] J.Maddox, Nature \textbf{361}, 397 (1993).

[2] B.P.Barber and S.J.Putterman, Phys. Rev. Lett. \textbf{69}, 3839 (1992).

[3] R.Hiller, S.J.Putterman, and B.P.Barber, Phys. Rev. Lett. \textbf{69},
1182 (1992).

[4] B.P.Barber, S.J.Putterman, Nature \textbf{352}, 318 (1992).

[5] L.Yuan, Phys. Rev. \textbf{E} \textbf{72}, 046309 (2005).

[6] F.V.Prigara, E-print archives, astro-ph/0110483.

[7] F.V.Prigara, E-print archives, quant-ph/0501103.

[8] M.J.Moran, R.E.Haigh, M.E.Lowry et al., Nucl. Instr. Meth. \textbf{B}
\textbf{96}, 651 (1995).

[9] R.Hiller, K.Weninger, S.J.Putterman, B.P.Barber, Science \textbf{266},
248 (1994).

[10] M.Kuraica and N.Konjevic, Phys. Rev. \textbf{A} \textbf{46}, 4429
(1992).

\end{document}